\documentclass[pra,showpacs,amsmath,amssymb]{revtex4}
\usepackage{bm}
\usepackage{amsmath,amssymb}
\begin{document}

\title{Necessity of macroscopic operation for
the creation of superpositions
of macroscopically distinct states}

\author{Tomoyuki Morimae}
\email{morimae@asone.c.u-tokyo.ac.jp}
\affiliation{
Department of Basic Science, University of Tokyo, 
3-8-1 Komaba, Tokyo 153-8902, Japan}
\altaffiliation[Present address:]{
Energy and Environmental Systems Laboratory,
Hitachi, Ltd., 
7-2-1 Omika-cho, Hitachi city, Ibaraki 319-1221, Japan
}
\date{\today}
            
\begin{abstract}
We consider the creation of superpositions of macroscopically
distinct states by a completely-positive (CP) operation on a subsystem.
We conclude that the subsystem on which the CP operation acts
must be macroscopically large
if the success probability of the CP operation does
not vanish in the
thermodynamic limit.
In order to obtain this conclusion, we show
two inequalities each of which represents
a trade-off relation among the magnitude of an indicator
for superpositions of macroscopically distinct states,
the success probability of a CP operation, and the volume
of the subsystem on which the CP operation acts.
\end{abstract}
\pacs{03.65.-w, 03.65.Ta, 03.67.-a}
\maketitle  
\section{Introduction}
The change of a physical properties 
of the total system
under a certain operation on a subsystem has long been
attracting much special attentions in quantum physics.
For example, in quantum measurement theory~\cite{Neumann,braginsky},
one of the most important quantities
is the change of the probability distribution of an observable
under a projective measurement of another observable on the apparatus.
Recently, the change of the magnitude of
entanglement under local operations 
on subsystems and classical communications
among them
has also been intensively studied
in quantum information and quantum computation~\cite{Nielsen,bennett}.

In this paper, we consider the change of the
magnitude of two indicators, 
$p$~\cite{SM} and $q$~\cite{SM05},
for superpositions of macroscopically distinct states
under a completely-positive (CP) 
operation on a subsystem,
in order to quantitatively investigate the creation of superpositions of 
macroscopically distinct states~\cite{schrodinger,leggett,mermin} by such an operation.
In other words, 
we first prepare a state which does not contain a superposition
of macroscopically distinct states, and next perform a CP operation
on a subsystem so that the state changes into another state
which contains a superposition of macroscopically
distinct states.
The existence of a superposition of macroscopically
distinct states, if any,
is identified by calculating the indicators $p$ or $q$.
By deriving two trade-off relations among 
the magnitude of the indicators, the success probability
of a CP operation, and the volume of the subsystem
on which the CP operation acts,
we generally show that if the success probability of the CP operation
does not vanish in the thermodynamic limit, the subsystem
on which the CP operation acts
must be macroscopically large, i.e., of the order of the size
of the total system.
This means {\it the necessity of a macroscopic operation for the creation
of a superposition of macroscopically distinct states.}

The outline of this paper is as follows.
Firstly, we consider a CP operation which maps a pure state to
a pure state.
As the indicator for superposition of macroscopically distinct states
in pure states, we use the index $p$~\cite{SM,Morimae,visual,Sugita},
which will be briefly reviewed in the next section for the convenience
of the reader.
In Sec.~\ref{section:p}, we will derive a trade-off relation among
the magnitude of the index $p$, the success probability of a CP operation,
and the volume of the subsystem on which the CP operation acts.
From this trade-off relation, we conclude that
it is necessary to access a macroscopically large subsystem
to create a superposition of macroscopically distinct states
by a CP operation with a non-vanishing success probability
in the thermodynamic limit.

Secondly, we consider more general CP operation which maps a mixed state
to a mixed state.
In this case, we use another indicator for superpositions of macroscopically
distinct states, which is called the index $q$~\cite{SM05},
since the index $p$ is not an indicator for 
superpositions of macroscopically
distinct states if the state is mixed.
After briefly explaining the definition of the index $q$ 
in Sec.~\ref{section:defq},
we derive a similar trade-off relation as that for the index $p$
in Sec.~\ref{section:q}.
This trade-off relation again means the necessity of
a macroscopic operation for 
the creation of a superposition of macroscopically
distinct states with a non-vanishing success probability
in the thermodynamic limit.

\section{Pure states: index $p$}
As the indicator for superpositions of macroscopically
distinct states in pure states, we use the index $p$.
For the convenience of the reader, we here briefly review
the definition of the index $p$. 
For more details, see Refs.~\cite{SM,visual,Morimae,Sugita}.

Let us consider an $N$-site lattice,
where $N$ is large but finite ($1\ll N<\infty$).
For example, a ring of $N$ spin-$1/2$ particles,
which is often studied in condensed matter
physics,
is one example of such systems.
Throughout this paper, 
we assume that the 
dimension of the Hilbert space on each site is 
an $N$-independent constant.

We use two symbols, $O$ and $o$, in order to represent 
asymptotic behaviors of a function $f(N)$
in the thermodynamic limit $N \to \infty$.
Firstly,
$f(N)=O(N^n)$ means
\begin{eqnarray*}
\lim_{N\to\infty}\frac{f(N)}{N^n}=\mbox{const.}\neq0.
\end{eqnarray*}
Secondly, $f(N)=o(N^n)$ means
\begin{eqnarray*}
\lim_{N\to\infty}\frac{f(N)}{N^n}=0.
\end{eqnarray*}
For example, if $f(N)=3N^5+2N^2+9$, we denote
$f(N)=O(N^5)$ or $f(N)=o(N^6)$.

For a given pure state $|\psi\rangle$, 
the index $p$ $(1\leq p \leq 2)$ is defined by
\begin{eqnarray*}
\max_{\hat{A}}
\Big[\langle\psi|\hat{A}^2|\psi\rangle
-\langle\psi|\hat{A}|\psi\rangle^2\Big]
=O(N^p),
\label{defp}
\end{eqnarray*}
where 
the maximum is taken over all Hermitian additive operators 
$\hat{A}$.
Here, an additive operator~\cite{grib,poulin} 
\begin{eqnarray*}
\hat{A}
=\sum_{l=1}^N\hat{a}(l)
\end{eqnarray*}
is a sum of local operators $\{\hat{a}(l)\}_{l=1}^N$,
where 
\begin{eqnarray*}
\hat{a}(l)\equiv
\hat{1}(1)\otimes\cdots\otimes\hat{1}(l-1)\otimes
\hat{a}(l)
\otimes\hat{1}(l+1)\otimes
\cdots\otimes\hat{1}(N)
\end{eqnarray*}
is a local operator acting on site $l$.
For example, if the system is a ring 
of $N$ spin-$1/2$ particles,
$\hat{a}(l)$ is a linear combination
of three Pauli operators, 
$\hat{\sigma}_x(l),\hat{\sigma}_y(l),\hat{\sigma}_z(l)$, 
and the identity operator $\hat{1}(l)$.
In this case, the $x$-component of the total
magnetization 
\begin{eqnarray*}
\hat{M}_x\equiv\sum_{l=1}^N\hat{\sigma}_x(l)
\end{eqnarray*}
and the $z$-component of the total staggard magnetization
\begin{eqnarray*}
\hat{M}_z^{st}\equiv\sum_{l=1}^N(-1)^l\hat{\sigma}_z(l)
\end{eqnarray*}
are, for example, additive operators.
In order to make $\hat{A}$ additive,
we assume that each $\hat{a}(l)$ is independent of $N$.
Since multiplying $\hat{a}(l)$ by an $N$-independent constant
does not change the essential results of this paper,
we henceforth assume that $\hat{a}(l)$ is normalized as 
$\|\hat{a}(l)\|_\infty\le 1$
without loss of generality, 
where $\|\hat{X}\|_\infty$ is the operator norm of $\hat{X}$.

The index $p$ takes the minimum value 1 for any 
``product state"
\begin{eqnarray*}
|\psi\rangle=\bigotimes_{l=1}^N|\phi_l\rangle,
\end{eqnarray*}
where $|\phi_l\rangle$ is a state of site $l$.
If $p$ takes the maximum value 2, on the other hand,
$|\psi\rangle$ contains a superposition of macroscopically 
distinct states,
because in this case a Hermitian additive operator  
has a ``macroscopically large" fluctuation
in the sense that
\begin{eqnarray*}
\lim_{N\to\infty}\frac{\sqrt{
\langle\psi|\hat{A}^2|\psi\rangle
-\langle\psi|\hat{A}|\psi\rangle^2
}}{N}
\neq0,
\end{eqnarray*}
and because the fluctuation of an observable
in a {\it pure} state means the existence of a superposition
of eigenvectors of the observable
corresponding to different eigenvalues~\cite{Morimae,visual}.
Here, we say that two eigenvectors
$|A_1\rangle$ and
$|A_2\rangle$
of an additive operator $\hat{A}$
corresponding to eigenvalues $A_1$ and $A_2$, respectively,
are macroscopically distinct with each other
if $A_1-A_2=O(N)$.

For example, the ``Cat state" 
\begin{eqnarray*}
|C\rangle\equiv
\frac{1}{\sqrt{2}}
\big(|0^{\otimes N}\rangle
+|1^{\otimes N}\rangle\big)
\end{eqnarray*}
has $p=2$, since
\begin{eqnarray*}
\langle C|\hat{M}_z^2|C\rangle
-\langle C|\hat{M}_z|C\rangle^2=O(N^2),
\end{eqnarray*}
and therefore contains 
a superposition of macroscopically
distinct states.

For the derivation of the trade-off relation in the next section,
it is useful to point out  
another representation of the definition of the index $p$.
Let
\begin{eqnarray*}
\big\|\hat{X}\big\|_k\equiv\Big(\sum_{j=1}^r|e_j|^k\Big)^{1/k}
\end{eqnarray*}
be the $k$-norm of an operator $\hat{X}$, where
$e_j$ is $j$-th eigenvalue of $\hat{X}$ and $r$ is the
rank of $\hat{X}$~\cite{matrix}.
Then, as will be shown below, the relation
\begin{eqnarray}
\max_{\hat{A}}\Big\|\big[\hat{A},|\psi\rangle\langle\psi|\big]\Big\|_k
=O(N^{\frac{p}{2}})
\label{noncomm}
\end{eqnarray}
holds for any $k$,
where 
the maximum is taken over all Hermitian additive operators $\hat{A}$.
This equation gives another physical meaning of the index $p$:
the index $p$ quantifies the ``noncommutativity" between the state
and an additive operator~\cite{janzing}.
In other words, {\it
a superposition of macroscopically distinct states
can also be detected by finding a ``macroscopically large"
noncommutativity.}

In particular, if we consider the operator norm, i.e., $k=\infty$, 
\begin{eqnarray*}
\max_{\hat{A}}\Big\|\big[\hat{A},|\psi\rangle\langle\psi|\big]\Big\|_\infty
=O(N^{\frac{p}{2}})
\end{eqnarray*}
is satisfied. This relation will be used in the next section
in order to derive the trade-off relation.

{\it Proof of Eq.~(\ref{noncomm})}:
Let us rewrite $\hat{A}|\psi\rangle$
as
\begin{eqnarray*}
\hat{A}|\psi\rangle=
\langle\psi|\hat{A}|\psi\rangle|\psi\rangle
+
\sqrt{\langle\psi|\hat{A}^2|\psi\rangle
-\langle\psi|\hat{A}|\psi\rangle^2}
|\phi\rangle,
\end{eqnarray*}
where
\begin{eqnarray*}
|\phi\rangle\equiv
\frac{\hat{A}-\langle\psi|\hat{A}|\psi\rangle}
{\sqrt{\langle\psi|\hat{A}^2|\psi\rangle
-\langle\psi|\hat{A}|\psi\rangle^2}}
|\psi\rangle
\end{eqnarray*}
is a normalized vector
orthogonal to $|\psi\rangle$.
Then, we obtain
\begin{eqnarray*}
i\big[\hat{A},|\psi\rangle\langle\psi|\big]
=
i
\sqrt{
\langle\psi|\hat{A}^2|\psi\rangle
-\langle\psi|\hat{A}|\psi\rangle^2}
\Big(|\phi\rangle\langle\psi|
-|\psi\rangle\langle\phi|\Big),
\end{eqnarray*}
which means that the eigenvalues of the Hermitian operator
$i\big[\hat{A},|\psi\rangle\langle\psi|\big]$
are 
\begin{eqnarray*}
\pm\sqrt{
\langle\psi|\hat{A}^2|\psi\rangle
-\langle\psi|\hat{A}|\psi\rangle^2}.
\end{eqnarray*}
Hence Eq.~(\ref{noncomm}) has been shown.

\section{Trade-off relation for pure states}
\label{section:p}
Let us consider a CP operation on a subsystem $S$
which maps a pure state $|\psi_1\rangle$ 
to a pure state $|\psi_2\rangle$.
By considering the normalization of the 
state after the CP operation, 
the state change is generally 
written as
\begin{eqnarray*}
|\psi_1\rangle\to
|\psi_2\rangle
\equiv\frac{\hat{E}|\psi_1\rangle}
{\sqrt{\langle\psi_1|\hat{E}^\dagger\hat{E}|\psi_1\rangle}},
\end{eqnarray*}
where $\hat{E}$ is a Kraus operator acting on $S$~\cite{Nielsen}.
For example, if the total system is 
a ring of $N$ spin-$1/2$ particles 
and we consider the spin-flip operation on
even sites,
the subsystem $S$ is the set of even sites and
the Kraus operator acting on $S$ is
\begin{eqnarray*}
\hat{E}=\prod_{l=even}\hat{\sigma}_x(l).
\end{eqnarray*}

Since 
$\langle\psi|\hat{E}^\dagger\hat{E}|\psi\rangle\le1$
for any state $|\psi\rangle$,
we obtain
$\|\hat{E}\|_\infty\le 1$.
By using the triangle inequality of the operator norm and
the fact that
\begin{eqnarray*}
\Big\|\big[\hat{A},\hat{E}\big]\Big\|_\infty
=\Big\|\sum_{l\in S}[\hat{a}(l),\hat{E}]\Big\|_\infty
\le2|S|,
\end{eqnarray*}
where $|S|$ is the volume of $S$ (i.e., the 
number of sites belonging to the subsystem $S$),
we obtain
\begin{eqnarray*}
\Big\|\big[\hat{A},|\psi_2\rangle\langle\psi_2|\big]\Big\|_\infty
&=&
\frac{1}{G}
\Big\|\big[\hat{A},
\hat{E}|\psi_1\rangle\langle\psi_1|\hat{E}^\dagger
\big]\Big\|_\infty\\
&=&
\frac{1}{G}
\Big\|
\big[\hat{A},\hat{E}\big]|\psi_1\rangle\langle\psi_1|\hat{E}^\dagger
+\hat{E}\big[\hat{A},|\psi_1\rangle\langle\psi_1|\big]\hat{E}^\dagger
+\hat{E}|\psi_1\rangle\langle\psi_1|\big[\hat{A},\hat{E}^\dagger\big]
\Big\|_\infty\\
&\le&
\frac{1}{G}
\Big(
\Big\|\big[\hat{A},\hat{E}\big]\Big\|_\infty
+\Big\|\big[\hat{A},|\psi_1\rangle\langle\psi_1|\big]\Big\|_\infty
+\Big\|\big[\hat{A},\hat{E}^\dagger\big]\Big\|_\infty
\Big)\\
&\le&
\frac{1}{G}
\Big(4|S|
+\Big\|\big[\hat{A},|\psi_1\rangle\langle\psi_1|\big]\Big\|_\infty
\Big)
\end{eqnarray*}
for any additive operator $\hat{A}$,
where
$G\equiv\langle\psi_1|\hat{E}^\dagger\hat{E}|\psi_1\rangle$
is the success probability of the CP operation.

Let us assume that the value of $p$ for $|\psi_1\rangle$ 
is $p_1$ and that for $|\psi_2\rangle$
is $p_2$. Then, the above inequality gives
\begin{eqnarray}
O(N^{p_2/2})
&\le&
\frac{1}{G}
\Big(4|S|
+O(N^{p_1/2})
\Big),
\label{result1}
\end{eqnarray}
which represents the trade-off relation among
the magnitude of the index $p$, the success probability $G$ of the CP operation,
and the volume $|S|$ of the subsystem $S$ on which the CP operation acts.

In the following two subsections, 
let us investigate this trade-off relation 
for the most important case 
$p_1<p_2=2$, i.e.,
the case where
a state which contains a superposition of macroscopically distinct states 
is created
from a state which does not contain 
a superposition of macroscopically distinct states.

\subsection{If $G=O(N^0)$}
Firstly, 
if the success probability of the
CP operation does not vanish 
in the thermodynamic limit, i.e., $G=O(N^0)$, 
Eq.~(\ref{result1}) gives
\begin{eqnarray*}
|S|=O(N),
\end{eqnarray*}
which means that {\it it is necessary to 
access a macroscopically large subsystem
in order to create a superposition of macroscopically distinct states 
from a state without such superpositions by 
a pure-to-pure CP operation with
a non-vanishing success probability as $N\to\infty$.}
This is one of the two main results of this paper 
(the other is given in Sec.~\ref{section:q}).

Inversely, $|S|=O(N)$ is also a sufficient condition
for the creation of a superposition of macroscopically distinct states
with a non-vanishing success probability as $N\to\infty$.
For example, 
it is easy to verify that
the state
\begin{eqnarray*}
|\psi_2\rangle=\frac{1}{\sqrt{2}}\Big(|0^{\otimes |S|}\rangle
+|1^{\otimes |S|}\rangle\Big)\otimes|\phi\rangle,
\end{eqnarray*}
which obviously has $p=2$ if $|S|=O(N)$,
can be created 
from any state $|\psi_1\rangle$ by a deterministic
CP operation (i.e., $G=1$) in the following manner.
Let
\begin{eqnarray*}
|\psi_1\rangle=\sum_{i=1}^v\lambda_i|\xi_i\rangle\otimes|\phi_i\rangle
\end{eqnarray*}
be a Schmidt decomposition~\cite{Nielsen} of the given state $|\psi_1\rangle$,
where $v$ is the Schmidt rank, $\sum_{i=1}^v|\lambda_i|^2=1$,
$|\xi_i\rangle$'s are states of the subsystem $S$,
and $|\phi_i\rangle$'s are states of other sites.
Then, the application of the operator
\begin{eqnarray*}
\frac{1}{\sqrt{2}}\Big(|0^{\otimes |S|}\rangle
+|1^{\otimes |S|}\rangle\Big)\sum_{i=1}^v\langle\xi_i|,
\end{eqnarray*}
which acts on the subsystem $S$,
to the state $|\psi_1\rangle$
deterministically creates $|\psi_2\rangle$.

\subsection{If $G=o(N^0)$}
Secondly, if $G=o(N^0)$, 
an access to a macroscopically large subsystem
is no longer necessary.
For example, let us
consider the state 
\begin{eqnarray*}
\frac{1}{N^\alpha}|1^{\otimes N}\rangle
+\sqrt{1-\frac{1}{N^{2\alpha}}}|0^{\otimes N}\rangle
\end{eqnarray*}
with $0<\alpha \le\frac{1}{2}$.
By using the VCM method~\cite{Morimae,visual,Sugita},
which is a method of efficiently calculating the 
magnitude of the index $p$,
it is straightforward to show that
this state has $p=2-2\alpha$,
and therefore does not contain a superposition
of macroscopically distinct states.
However, this state
can be changed into the state
\begin{eqnarray*}
|\phi\rangle\otimes
\frac{1}{\sqrt{2}}\Big(|0^{\otimes N-1}\rangle
+|1^{\otimes N-1}\rangle\Big),
\end{eqnarray*}
which obviously has $p=2$,
by the local projection 
\begin{eqnarray*}
\Big(\sqrt{1-\frac{1}{N^{2\alpha}}}|1\rangle+\frac{1}{N^\alpha}|0\rangle\Big)
\Big(\sqrt{1-\frac{1}{N^{2\alpha}}}\langle1|+\frac{1}{N^\alpha}\langle0|\Big)
\end{eqnarray*}
on the single site
with the success probability
$G=O(N^{-2\alpha})$.
Here, $|\phi\rangle$ is a state of a single site.
Indeed, in this case,  
the right-hand side of
the inequality~(\ref{result1}) is 
$O(N^{1+\alpha})$,
and therefore the inequality
is not violated.

\section{Mixed states: index $q$}
\label{section:defq}
In the previous sections, 
we have studied the change of the magnitude of
the index $p$
under a pure-to-pure CP operation.
If we consider {\it mixed} states, however, the naive generalization
of the index $p$ to mixed states: 
\begin{eqnarray*}
\max_{\hat{A}}\Big[
\mbox{Tr}(\hat{\rho}\hat{A}^2)
-\mbox{Tr}(\hat{\rho}\hat{A})^2
\Big]=O(N^p)
\end{eqnarray*}
is no longer 
a good indicator for superpositions of macroscopically distinct states,
since a fluctuation is not necessarily equivalent to a superposition
if the state is mixed.
For example, let us consider a ``classical mixture" of 
two macroscopically distinct states:
\begin{eqnarray*}
\hat{\rho}=\frac{1}{2}|0^{\otimes N}\rangle\langle0^{\otimes N}|
+\frac{1}{2}|1^{\otimes N}\rangle\langle1^{\otimes N}|.
\end{eqnarray*}
Although it is obvious that this state contains no superposition
of macroscopically distinct states,
this state has $p=2$
in terms of the above generalization
since
\begin{eqnarray*}
\mbox{Tr}(\hat{\rho}\hat{M}_z^2)
-\mbox{Tr}(\hat{\rho}\hat{M}_z)^2=O(N^2).
\end{eqnarray*}
This example shows that we must use another 
indicator if the state is mixed.

A good indicator, the index $q$, 
for superpositions of macroscopically distinct states
in mixed states
was recently proposed in Ref.~\cite{SM05}.
For a given state $\hat{\rho}$, the index $q$ ($1\le q\le2$) is defined by
\begin{eqnarray*}
\max\Big(N,\max_{\hat{A},\hat{\eta}}\mbox{Tr}
\Big(\hat{\rho}[\hat{A},[\hat{A},\hat{\eta}]]\Big)\Big)
=O(N^q),
\end{eqnarray*}
where $\hat{\eta}$ is a projection operator 
and $\hat{A}$ is a Hermitian additive
operator.
As detailed in Ref.~\cite{SM05}, 
$q$ takes the minimum value 1 for any mixture
of product states:
\begin{eqnarray*}
\sum_i\lambda_i\Big(\bigotimes_{l=1}^N|\phi_l^i\rangle\Big)
\Big(\bigotimes_{l=1}^N\langle\phi_l^i|\Big),
\end{eqnarray*}
where $0\le\lambda_i\le1$, $\sum_i\lambda_i=1$,
and $|\phi_l^i\rangle$ is a state of site $l$.
On the other hand,
if $q$ takes the maximum value 2,
$\hat{\rho}$ contains a superposition
of macroscopically distinct states~\cite{macro}.

For later convenience, let us give another representation
of the definition of the index $q$.
Let $\hat{X}$ be any traceless operator.
Then,
\begin{eqnarray*}
\|\hat{X}\|_1=2\max_{\hat{\eta}}\mbox{Tr}(\hat{\eta}\hat{X})
\end{eqnarray*}
holds, 
where $\hat{\eta}$ is a projection operator~\cite{matrix}.
By using this relation
and the fact 
\begin{eqnarray*}
\mbox{Tr}\Big(\hat{\rho}[\hat{A},[\hat{A},\hat{\eta}]]\Big)
=\mbox{Tr}\Big(\hat{\eta}[\hat{A},[\hat{A},\hat{\rho}]]\Big),
\end{eqnarray*}
it is easy to verify that the index $q$ is also defined by
\begin{eqnarray*}
\max\Big(N,\max_{\hat{A}}
\Big\|[\hat{A},[\hat{A},\hat{\rho}]]\Big\|_1\Big)
=O(N^q).
\end{eqnarray*}
This expression gives a clear physical meaning of the index $q$:
{\it the index $q$ quantifies the ``noncommutativity" between the state
and an additive operator}.
\section{Trade-off relation for mixed states}
\label{section:q}
Let us 
consider a CP operation on a subsystem $S$
which maps a state $\hat{\rho}_1$ to a state $\hat{\rho}_2$.
By considering the normalization of the state
after the operation, the state change is generally written as
\begin{eqnarray*}
\hat{\rho}_1\to\hat{\rho}_2
\equiv\frac{
\sum_{k=1}^M \hat{E}_k\hat{\rho}_1 \hat{E}_k^\dagger}
{\mbox{Tr}\big(\sum_{k=1}^M \hat{E}_k\hat{\rho}_1 \hat{E}_k^\dagger\big)},
\end{eqnarray*}
where $\hat{E}_k$ is a Kraus operator acting on $S$~\cite{Nielsen}.

As is shown in Appendix~\ref{proofq},  
the inequality
\begin{eqnarray*}
\Big\|[\hat{A},[\hat{A},\hat{\rho}_2]]\Big\|_1
&\le&
\frac{1}{G}
\Big(
\Big\|\big[\hat{A},\big[\hat{A},\hat{\rho}_1\big]\big]\Big\|_1
+16|S|N
+4|S|^2G
+12|S|^2
\Big)
\end{eqnarray*}
holds for any additive operator $\hat{A}$,
where
\begin{eqnarray*}
G\equiv\mbox{Tr}\Big(\sum_{k=1}^M\hat{E}_k^\dagger\hat{E}_k\hat{\rho}_1\Big)
\end{eqnarray*}
is the success probability of the CP operation.

As in the case of the index $p$, this
inequality means the trade-off relation among
the magnitude of the index $q$, the success
probability $G$ of the CP operation, and the volume $|S|$
of the subsystem $S$ on which the CP operation acts.
In particular, if we consider 
the creation of a state having $q=2$ from a state having 
$q<2$ by a CP operation with a non-vanishing success 
probability as $N\to\infty$ (i.e., $G=O(N^0)$),
this trade-off relation gives the same necessary condition:
\begin{eqnarray*}
|S|=O(N)
\end{eqnarray*}
for $S$.
In other words, {\it it is necessary to 
access a macroscopically large subsystem
in order to create a superposition of macroscopically distinct states
by a mix-to-mix CP operation
with a non-vanishing success probability as $N\to\infty$}.
This is the other main result of this paper.

\section{Conclusion}
In this paper, we have 
studied the change of the magnitude of
two indicators, $p$ and $q$, 
for superpositions
of macroscopically distinct states under a 
creation of such superpositions by a CP operation on a subsystem.
By deriving two trade-off relations among
the magnitude of the indicators, the success probability of a CP operation,
and the volume of the subsystem on which the CP operation acts,
we have generally shown that
it is necessary to access a macroscopically large subsystem
to create a superposition of macroscopically distinct states
by a CP operation with a non-vanishing success probability
in the thermodynamic limit.

\acknowledgements
The author thanks A. Shimizu and Y. Matsuzaki for useful discussions.
This work was partially supported by Japan Society for the
Promotion of Science.
\appendix*

\section{}
\label{proofq}
In this appendix, we will show the inequality given in Sec.~\ref{section:q}.
We assume that the dimension of the Hilbert space on each site
is an $N$-independent constant and $\|\hat{a}(l)\|_\infty\le 1$.

Let us first mention a useful lemma.

{\it Lemma}:
For any Hermitian operator $\hat{X}$,
\begin{eqnarray*}
\Big\|\sum_{k=1}^M\hat{E}_k\hat{X}\hat{E}_k^\dagger\Big\|_1
\le\|\hat{X}\|_1.
\end{eqnarray*}

{\it Proof}:
Since
\begin{eqnarray*}
\mbox{Tr}\Big(\sum_{k=1}^M\hat{E}_k^\dagger\hat{E}_k\hat{\rho}\Big)\le 1
\end{eqnarray*}
for any $\hat{\rho}$, we obtain 
\begin{eqnarray*}
\sum_{k=1}^M\hat{E}_k^\dagger\hat{E}_k\le \hat{1}.
\end{eqnarray*}
Let
\begin{eqnarray*}
\hat{X}=\sum_{i=1}^rx_i|i\rangle\langle i|
\end{eqnarray*}
be a spectral decomposition of $\hat{X}$.
Then,
\begin{eqnarray*}
\Big\|\sum_{k=1}^M\hat{E}_k\hat{X}\hat{E}_k^\dagger\Big\|_1
&\le&
\sum_{k=1}^M
\sum_{i=1}^r|x_i|~
\big\|
\hat{E}_k|i\rangle\langle i|\hat{E}_k^\dagger
\big\|_1\\
&=&
\sum_{k=1}^M
\sum_{i=1}^r|x_i|~
\langle i|\hat{E}_k^\dagger \hat{E}_k|i\rangle\\
&\le&\sum_{i=1}^r|x_i|\\
&=&\|\hat{X}\|_1.
\end{eqnarray*}
Hence the lemma has been shown.

By using this lemma, let us next show the inequality.
Consider
\begin{eqnarray*}
\Big\|\big[\hat{A},\big[\hat{A},\hat{\rho}_2\big]\big]\Big\|_1
&=&
\frac{1}{G}
\left\|
\sum_{k=1}^M
\left[\hat{A},
\left[\hat{A},
\hat{E}_k\hat{\rho}_1\hat{E}_k^\dagger
\right]
\right]
\right\|_1
\le
\frac{1}
{G}
\sum_{j=1}^3
\big\|\hat{\Xi}_j\big\|_1,
\end{eqnarray*}
where
\begin{eqnarray*}
\hat{\Xi}_1
&\equiv&
\sum_{k=1}^M
\hat{E}_k[\hat{A},[\hat{A},\hat{\rho}_1]]\hat{E}_k^\dagger,\\
\hat{\Xi}_2
&\equiv&
2\sum_{k=1}^M
\Big(
[\hat{A},\hat{E}_k][\hat{A},\hat{\rho}_1]\hat{E}_k^\dagger
+\hat{E}_k[\hat{A},\hat{\rho}_1][\hat{A},\hat{E}_k^\dagger]
\Big),\\
\hat{\Xi}_3
&\equiv&
\sum_{k=1}^M
\Big(
[\hat{A},[\hat{A},\hat{E}_k]]\hat{\rho}_1\hat{E}_k^\dagger
+\hat{E}_k\hat{\rho}_1[\hat{A},[\hat{A},\hat{E}_k^\dagger]]
+2[\hat{A},\hat{E}_k]\hat{\rho}_1[\hat{A},\hat{E}_k^\dagger]
\Big).
\end{eqnarray*}

Since $[\hat{A},[\hat{A},\hat{\rho}_1]]$ is Hermitian, 
\begin{eqnarray*}
\|\hat{\Xi}_1\|_1\le
\Big\|[\hat{A},[\hat{A},\hat{\rho}_1]]\Big\|_1
\end{eqnarray*}
from the lemma.

Let us define 
\begin{eqnarray*}
\hat{A}_S\equiv\sum_{l\in S}\hat{a}(l).
\end{eqnarray*}
Then,
\begin{eqnarray*}
\|\hat{\Xi}_2\|_1
&=&2
\Big\|
\sum_{k=1}^M
\Big(
[\hat{A}_S,\hat{E}_k][\hat{A},\hat{\rho}_1]\hat{E}_k^\dagger
+\hat{E}_k[\hat{A},\hat{\rho}_1][\hat{A}_S,\hat{E}_k^\dagger]
\Big)
\Big\|_1\\
&=&2
\Big\|
\big[\hat{A}_S,\sum_{k=1}^M\hat{E}_k[\hat{A},\hat{\rho}_1]\hat{E}_k^\dagger\big]
+\sum_{k=1}^M\hat{E}_k[[\hat{A},\hat{\rho}_1],\hat{A}_S]\hat{E}_k^\dagger
\Big\|_1\\
&\le&16|S|N
\end{eqnarray*}
and
\begin{eqnarray*}
\|\hat{\Xi}_3\|_1
&=&
\Big\|
\sum_{k=1}^M
\Big(
[\hat{A}_S,[\hat{A}_S,\hat{E}_k]]\hat{\rho}_1\hat{E}_k^\dagger
+\hat{E}_k\hat{\rho}_1[\hat{A}_S,[\hat{A}_S,\hat{E}_k^\dagger]]
+2[\hat{A}_S,\hat{E}_k]\hat{\rho}_1[\hat{A}_S,\hat{E}_k^\dagger]
\Big)
\Big\|_1\\
&=&
\Big\|
[\hat{A}_S,[\hat{A}_S,\sum_{k=1}^M\hat{E}_k\hat{\rho}_1\hat{E}_k^\dagger]]
+\sum_{k=1}^M
\hat{E}_k[\hat{A}_S,[\hat{A}_S,\hat{\rho}_1]]\hat{E}_k^\dagger
+2\big[
\hat{A}_S,\sum_{k=1}^M\hat{E}_k[\hat{\rho}_1,\hat{A}_S]\hat{E}_k^\dagger
\big]
\Big\|_1\\
&\le&
4|S|^2G
+12|S|^2.
\end{eqnarray*}
Therefore, we finally obtain
\begin{eqnarray*}
\Big\|\big[\hat{A},\big[\hat{A},\hat{\rho}_2\big]\big]\Big\|_1
\le
\frac{1}
{G}
\Big(
\Big\|[\hat{A},[\hat{A},\hat{\rho}_1]]\Big\|_1
+16|S|N
+4|S|^2G
+12|S|^2
\Big).
\end{eqnarray*}
Hence the inequality has been shown.



\begin{thebibliography}{00}
\bibitem{macro}
We say that two eigenvectors $|A_1\rangle$ and
$|A_2\rangle$ of an additive operator $\hat{A}$ corresponding
to eigenvalues $A_1$ and $A_2$, respectively, are macroscopically
distinct if $A_1-A_2=O(N)$.
As is seen in Eq.~(3) of Ref.~\cite{SM05}, the index $q$ is related 
to off-diagonal elements of the state under consideration.
Since off-diagonal elements represent quantum coherence,
the state contains a superposition of
macroscopically distinct states in the above sense if $q=2$.
\end{thebibliography}
\end{document}